\documentclass[11pt]{article}
\usepackage{enumerate}

\usepackage{amsmath}
\usepackage{float}
\usepackage{dsfont}
\usepackage{relsize}
\usepackage{amsfonts}
\usepackage{amssymb}
\usepackage{mathtools}
\usepackage[authoryear]{natbib}

\usepackage{amsthm}
\usepackage[toc,page]{appendix}

\usepackage{tikz}
\usetikzlibrary{shapes,arrows}
\newtheorem*{prop*}{Proposition}
\usepackage{fancyhdr}

\usepackage{fullpage}
\newcommand\independent{\protect\mathpalette{\protect\independenT}{\perp}}
\def\independenT#1#2{\mathrel{\rlap{$#1#2$}\mkern2mu{#1#2}}}

\begin{document}

\begin{center}
\textbf{Inverse Probability Weighted Estimators of Vaccine Effects \\
Accommodating Partial Interference and Censoring}

Sujatro Chakladar$^{1}$,  Michael G. Hudgens$^{1*}$, M. Elizabeth Halloran$^{2,3}$,\\ 
John D. Clemens$^4$, Mohammad Ali$^5$ and Michael E. Emch$^6$ 
\\
\end{center}

\noindent $^{1}$Department of Biostatistics, University of North Carolina, 
Chapel Hill, North Carolina, U.S.A.\\
$^{2}$Department of Biostatistics, University of Washington, Seattle, Washington, U.S.A.\\
$^{3}$Vaccine and Infectious Disease Division, Fred Hutchinson Cancer Research Center, Seattle, Washington, U.S.A.\\
$^{4}$Department of Epidemiology, University of California, Los Angeles, California, U.S.A.\\
$^{5}$Department of International Health, Johns Hopkins University, Baltimore, Maryland, U.S.A.\\
$^{6}$Department of Geography, University of North Carolina, Chapel Hill, North Carolina, U.S.A.\\
$*$ mhudgens@email.unc.edu\\

{\bf Abstract}: Estimating population-level effects of a vaccine is challenging because there may be interference, i.e.,
 the outcome of one individual may depend on the vaccination status of another individual. Partial interference occurs when individuals can be partitioned into groups such that interference occurs only within groups. In the absence of interference, inverse probability weighted (IPW) estimators are commonly used to draw inference about causal effects of an exposure or treatment. Tchetgen Tchetgen and VanderWeele (2012) proposed a modified IPW estimator for causal effects in the presence of partial interference. Motivated by a cholera vaccine study in Bangladesh,
 this paper considers an extension of the Tchetgen
Tchetgen and VanderWeele IPW estimator to the setting where the outcome is subject to right censoring using inverse probability of censoring weights (IPCW). Censoring weights are estimated using proportional hazards frailty models. The large sample properties of the IPCW estimators are derived, and simulation studies are presented demonstrating the estimators' performance in finite samples. The methods are then used to analyze data from the cholera vaccine study.

\newpage

\section{Introduction}

Estimating population-level effects of a vaccine is challenging because there may be interference, i.e.,
 the outcome of one individual may depend on the vaccination status of another individual \citep{cox1958,  halloran1991study}. Partial interference is a special case of interference where individuals can be partitioned into groups such that interference does not occur between individuals in different groups but may occur between individuals in the same group (Sobel 2006). Partial interference might be a reasonable assumption if groups of individuals are sufficiently separated geographically, socially, and/or temporally. For example, in an assessment of the effects of cholera vaccination in a study in Bangladesh,  Perez-Heydrich et al.\ (2014) assumed partial interference based on the spatial location of residences of study participants.
  Effects due to interference, also known as spillover effects or peer effects, are of interest in many areas, including criminology, 
  developmental psychology, econometrics, education, imaging, political science, social media and network analysis, sociology, and spatial analyses.
 
\par
Inferential methods about spillover effects have been developed for randomized experiments \citep{rosenbaum2007interference, hudgens2008toward, eckles2016estimating, baird2014designing}. However, in some settings it may not be feasible or ethical to randomize groups or individuals to different treatment or exposure conditions. In the observational setting, Tchetgen Tchetgen and VanderWeele (henceforth TV) (2012) proposed inverse probability weighted (IPW) estimators for different types of causal effects when there may be partial interference. Large sample properties of these IPW estimators were considered by Perez-Heydrich et al.\ (2014) and \cite{liu2016inverse}. 
While motivated by observational studies, 
these estimators may also be applied in cluster (group) randomized trials where partial interference is assumed and there is non-compliance, i.e., not all individuals receive the treatment assigned to their cluster. These estimators are also applicable to settings such as the cholera vaccine study where all individuals in the study were randomized but only a subset chose to participate in the trial.

\par In settings where the outcome of interest is a time to event, the outcome may be subject to right censoring due to study completion or participant drop-out. For example, in the Bangladesh cholera vaccine trial mentioned above, some study participants emigrated out of the field trial area and hence were lost to follow-up. 
In the absence of interference, censoring is often accommodated by using inverse probability of censoring weights along with inverse probability treatment weights \citep{robins2000correcting,cole2008constructing}. In this paper, an extension of the TV IPW estimators is considered for the setting where there may be partial interference and the outcome is subject to right censoring using inverse probability of censoring weights (IPCW). 
\par The outline of this paper is as follows. The proposed methods are developed in Section 2. In Section 3 simulation results are presented demonstrating the empirical performance of the proposed methods in finite sample settings. In Section 4 the methods are used to analyze the Bangladesh  cholera vaccine study. Section 5 concludes with a discussion.

\section{ Methods}

\subsection{Estimands}

Suppose data are observed from $m$ groups of individuals, with $n_i$ individuals per group for $i=1,\ldots,m$. Let $A_{ij} = 1$ if individual $j$ in group $i$ receives treatment (e.g., vaccine) and $A_{ij} = 0$ otherwise. Let $\mathbf{A}_{i} = (A_{i1},A_{i2},...,A_{in_i})$ and $\mathbf{A}_{i,-j} = (A_{i1},A_{i2},...,A_{ij-1},A_{ij+1}...,A_{in_i})$. Let $\mathbf a_i$ and $\mathbf a_{i,-j}$ denote possible realizations of $\mathbf A_i$ and $\mathbf A_{i,-j}$, and let $\mathcal{A}(n)$ denote the set of all possible $2^n$ treatments for a group size of  $n = 1,2,\hdots$. Assume partial interference and denote the potential time to event for individual $j$ in group $i$ if, possibly counter to fact, group $i$ receives treatment $\mathbf{a}_i$ by $T_{ij}(\mathbf{a}_i)$. The notation $T_{ij}(\mathbf{a}_i)$ reflects the partial interference assumption, i.e., the potential outcome of individual $j$ in group $i$ does not depend on the treatment of individuals outside group $i$. Below the notation $T_{ij}(a, \mathbf{a}_{i, -j})$ is sometimes used to make explicit the treatment for individual $j$ and the treatment for all other individuals in group $i$. Let $\mathbf{T}_{i}(.) = \{T_{ij}(\mathbf{a}_i):\mathbf{a}_i \in \mathcal{A}(n_i),j = 1,2,\cdots, n_i\}$ denote the set of all potential event times for individuals in group $i$. Suppose the event times are subject to right censoring, e.g., due to loss to follow-up or study completion. Let $C_{ij}$ denote the potential censoring times for individual $j$ in group $i$. Assume that treatment has no effect on the censoring times, i.e., $C_{ij}$ does not depend on $\mathbf{a}_i$. This assumption is reasonable for the cholera vaccine study because both the individuals in the study as well as the study investigators were blinded to treatment assignment.
Let $\Delta_{ij} =1$ if $T_{ij}(\mathbf{A}_i)\leq C_{ij}$ and $\Delta_{ij} =0$ otherwise, and let $X_{ij} = \min (T_{ij}(\mathbf{A}_i),C_{ij})$. Define $\mathbf{X}_i = (X_{i1},X_{i2},\cdots,X_{in_i})$ and $\boldsymbol{\Delta}_i = (\Delta_{i1}, \Delta_{i2}, \cdots,\Delta_{in_i})$. Denote by $\mathbf{L}_{ij}$ the vector of baseline covariates for subject $j$ in group $i$ and by $\mathbf{L}_{i}$ the matrix of baseline covariates for all subjects in group $i$, i.e., $\mathbf{L}_{i} = (\mathbf{L}_{i1},\mathbf{L}_{i2},\cdots,\mathbf{L}_{in_i})$. Assume that the $m$ groups are randomly sampled from an infinite superpopulation of groups such that the observed data are $m$ i.i.d. copies of $\mathbf{O}_i = (\mathbf{L}_i, \mathbf{A}_i,\mathbf{X}_i,\boldsymbol{\Delta}_i)$.

In the absence of interference, treatment effects are typically defined as contrasts in mean potential outcomes for different counterfactual scenarios, e.g., the average treatment effect is usually defined as the difference in the mean potential outcome had all individuals received treatment versus had no individuals received treatment. Similarly, in the setting where there is partial interference, causal effects may be defined as contrasts in mean potential outcomes for different counterfactual scenarios \citep{hong2006evaluating, sobel2006randomized, hudgens2008toward, tchetgen2010causal}. Here we consider counterfactual scenarios where the marginal probability that an individual receives treatment, $\Pr_{\alpha}(A_{ij} = 1)$, equals $\alpha$  for different values of $\alpha \in (0,1)$. The notation $\Pr_{\alpha}(\cdot)$ indicates that the probability corresponds to the distribution under the counterfactual scenario. Specifically, the Bernoulli treatment allocation strategy (or policy) described in TV is considered wherein individuals independently select treatment with probability $\alpha$. Let $\pi(\mathbf{a}_i,\alpha)$ denote the probability that group $i$ receives treatment $\mathbf{a}_i$ under Bernoulli allocation strategy $\alpha$. That is, $\pi(\mathbf{a}_i,\alpha) = \Pr_\alpha(\mathbf{A}_i=\mathbf{a}_i) = \prod_{k=1}^{n_i} \alpha^{a_{ik}}(1-\alpha)^{1-a_{ik}}$. Similarly let $\pi(\mathbf{a}_{i,-j},\alpha) = \Pr_{\alpha}(\mathbf{A}_{i,-j} = \mathbf{a}_{i,-j} |A_{ij} = a) = \prod_{k=1,k\neq j}^{n_i}\alpha^{a_{ik}}(1-\alpha)^{1-a_{ik}}$.
\par
The causal estimands of interest defined below are contrasts in the risk of having an event by time $t$ for different combinations of treatment $a$ and allocation strategies $\alpha$. To define these estimands, let
$$\bar{F}_{ij}(t,a,\alpha) = \sum_{\mathbf{a}_{i,-j}\in  \mathcal{A}(n_i-1)}I \{ T_{ij}(a,\mathbf{a}_{i,-j})\leq t\} \pi(\mathbf{a}_{i,-j},\alpha),$$
and
$$\bar{F}_{ij}(t,\alpha) = \sum_{\mathbf{a}_{i}\in  \mathcal{A}(n_i)}I \{ T_{ij}(\mathbf{a}_{i})\leq t\} \pi(\mathbf{a}_{i},\alpha).$$
In words, $\bar{F}_{ij}(t,a,\alpha)$ is the probability that individual $j$ in group $i$ will have an event by time $t$ when receiving treatment $a$ and the group adopts policy $\alpha$. Likewise, $\bar{F}_{ij}(t,\alpha)$ is the probability that individual $j$ in group $i$ will have an event by time $t$ when the group adopts allocation strategy $\alpha$.
Denote the group average risks by $\bar{F}_i(t,a,\alpha) = n_i^{-1}\sum_{j=1}^{n_i} \bar{F}_{ij}(t,a,\alpha)$ and $\bar{F}_i(t,\alpha) = n_i^{-1}\sum_{j=1}^{n_i} \bar{F}_{ij}(t,\alpha)$. Let $\mu (t,a,\alpha) = E_{\alpha}\{ \bar{F}_i(t,a,\alpha)\}$ and $\mu (t,\alpha) = E_{\alpha}\{ \bar{F}_i(t,\alpha)\}$ where $E_{\alpha}\{.\}$ denotes the expected value under the counterfactual setting when policy $\alpha$ is adopted in the superpopulation of groups. In the cholera vaccine study described in Section $4$, $\mu (t,a,\alpha)$ denotes the average risk of acquiring cholera by time $t$ when an individual receives treatment $a$ and other individuals receive vaccine with probability $\alpha$. 

Various effects of treatment can be defined by contrasts in $\mu(t,a,\alpha)$ and $\mu(t,\alpha)$
\citep{tchetgen2010causal,perez2014assessing}. 
The direct effect is obtained by comparing the probability of an event when an individual receives treatment versus when not receiving treatment for a fixed allocation strategy. In particular, the direct effect at time $t$ corresponding to policy $\alpha$ is defined to be $DE(t,\alpha) = \mu(t,0,\alpha) - \mu(t,1,\alpha)$. The indirect (or spillover) effect is the difference in the probability of an event by time $t$ for two different policies when the individual does not receive treatment. Specifically, the indirect effect is given by $IE(t,\alpha_1, \alpha_2) = \mu(t,0,\alpha_1) - \mu(t,0,\alpha_2)$ for allocation strategies $\alpha_1$ and $\alpha_2$. An indirect effect can analogously be defined when an individual is vaccinated. The total effect is defined as the difference between the probability of an event by time $t$ when an individual does not receive treatment under policy $\alpha_1$ and when an individual receives treatment under policy $\alpha_2$, i.e., $TE(t,\alpha_1, \alpha_2) = \mu(t,0,\alpha_1) - \mu(t,1,\alpha_2)$. Finally, the overall effect is the difference between the probability of an event by time $t$ for policy $\alpha_1$ versus $\alpha_2$, i.e., $OE(t,\alpha_1, \alpha_2) = \mu(t,\alpha_1) - \mu(t,\alpha_2)$.

\subsection{Assumptions}

Assume the following:
\begin{enumerate}[I)]
\item Conditional independent treatment: $\mathbf{A}_i\independent \mathbf{T}_{i}(.) \mid \mathbf{L}_i$ 
\item Treatment positivity: $\Pr(\mathbf{A}_i=\mathbf{a}_i \mid \mathbf{L}_i)>0$ for all $\mathbf{a}_i \in \mathcal{A}(n_i)$
\item Conditional independent censoring: $C_{ij} \independent T_{ij}(\mathbf{A}_{i})\mid \{\mathbf{L}_{i}, \mathbf{A}_i\}$
\item Non-censoring positivity: $\Pr(\Delta_{ij}=1 \mid \mathbf{L}_i , \mathbf{A}_i )>0$
\end{enumerate}
\par Assumption $\mathrm{I}$ states that the potential event times for individuals within the same group are conditionally independent of the actual treatment received by the group given covariates; this is a group-level generalization of the usual individual-level no unmeasured confounders assumption often made in the absence of interference \citep{tchetgen2010causal}.
Assumption I would be violated if there was some common cause of one or more components of $\mathbf{A}_i$ and $\mathbf{T}_i(\cdot)$ not included in $\mathbf{L}_{i}$.
Treatment positivity assumes that each group has a non-zero probability of being assigned every possible treatment combination given covariates for the group \citep{perez2014assessing}. 
Assumption II would not hold if there was some group which had zero chance of receiving some treatment combination (e.g., treatment $a=1$ for all individuals in the group).
Assumption $\mathrm{III}$ supposes that conditional on baseline group covariates and group treatment, an individual's failure time is independent of their censoring time.
Assumption III would be violated if there was some variable not in $\mathbf{L}_i$ or $\mathbf{A}_i$  which was prognostic of both the censoring and failure times. 
 Finally Assumption $\mathrm{IV}$ indicates that each individual has a non-zero probability of not being censored at each observation time \citep{rotnitzky2007}. 
In the next section 
  IPW estimators are proposed and shown to be consistent (and asymptotically normal)  for the direct, indirect, total, and overall effects under Assumptions I-IV.

\subsection{Proposed Estimator} 

In the absence of censoring, the IPW estimator proposed by TV can be used to draw inference about 
$\mu(t, a, \alpha)$ and $\mu(t, \alpha)$, i.e., the mean potential outcomes under the counterfactual setting where policy $\alpha$ is adopted.
In particular, letting $Y_{ij} = I(X_{ij} \leq t)$ be the indicator variable that the observation time for individual $j$ in group $i$ is less than or equal to $t$, the TV IPW estimators are $\hat{\mu}^{TV}(t,a,\alpha) = m^{-1}\sum_{i=1}^{m}{\hat{F}^{TV}_i(t,a,\alpha)}$ and $\hat{\mu}^{TV}(t,\alpha) = m^{-1}\sum_{i=1}^{m}{\hat{F}^{TV}_i(t,\alpha)}$ where
$$\hat{F}^{TV}_i(t,a,\alpha) = n_i^{-1}\sum_{j=1}^{n_i}\frac{\pi (\mathbf{A}_{i,-j};\alpha)I(A_{ij}=a)Y_{ij}}{\Pr (\mathbf{A}_i|\mathbf{L}_i,\hat{\boldsymbol{\beta}})}, \qquad \hat{F}^{TV}_i(t,\alpha) = n_i^{-1}\sum_{j=1}^{n_i}\frac{\pi (\mathbf{A}_{i};\alpha)Y_{ij}}{\Pr (\mathbf{A}_i|\mathbf{L}_i,\hat{\boldsymbol{\beta}})}, $$
and $\hat{\boldsymbol{\beta}}$ is an estimator of the vector of parameters for the propensity model $\Pr (\mathbf{A}_i|\mathbf{L}_i,\boldsymbol{\beta})$. Details of the propensity model are discussed in the next sections. 
\par In the presence of censoring, the following extension of the TV IPW estimators is proposed: $\hat{\mu}(t,a,\alpha) = m^{-1}\sum_{i=1}^{m}{\hat{F}_i(t,a,\alpha)}$ and $\hat{\mu}(t,\alpha) = m^{-1}\sum_{i=1}^{m}{\hat{F}_i(t,\alpha)}$
where
$$\hat{F}_i(t,a,\alpha) = n_i^{-1}\sum_{j=1}^{n_i}\frac{\pi (\mathbf{A}_{i,-j};\alpha)I(A_{ij}=a)I(\Delta_{ij}=1)I(X_{ij}\leq t)}{\Pr (\mathbf{A}_i|\mathbf{L}_i,\hat{\boldsymbol{\beta}})
S_C (X_{ij}|\mathbf{L}_{i},\mathbf{A}_i,\hat{\boldsymbol{\gamma}} )}, $$
$$\hat{F}_i(t,\alpha) = n_i^{-1}\sum_{j=1}^{n_i}\frac{\pi (\mathbf{A}_{i};\alpha)I(\Delta_{ij}=1)I(X_{ij}\leq t)}{\Pr (\mathbf{A}_i|\mathbf{L}_i,\hat{\boldsymbol{\beta}})
S_C (X_{ij}|\mathbf{L}_{i},\mathbf{A}_i, \hat{\boldsymbol{\gamma}})},$$
$S_C(t|\mathbf{L}_{i},\mathbf{A}_i, \hat{\boldsymbol{\gamma}})= \Pr(C_{ij}>t \mid \mathbf{L}_{i},\mathbf{A}_i, \hat{\boldsymbol{\gamma}})$
and $\hat{\boldsymbol{\gamma}}$ is an estimator of the vector of the parameters for the censoring model.  Details of the censoring model are discussed in the next sections. Estimates of the direct, indirect, total, and overall effects are given by $ \widehat{DE}(t,\alpha) = \hat{\mu}(t,0,\alpha) - \hat{\mu}(t,1,\alpha)$, $ \widehat{IE}(t,\alpha_1, \alpha_2) = \hat{\mu}(t,0,\alpha_1) - \hat{\mu}(t,0,\alpha_2)$, $ \widehat{TE}(t,\alpha_1, \alpha_2) = \hat{\mu}(t,0,\alpha_1) - \hat{\mu}(t,1,\alpha_2)$ and $ \widehat{OE}(t,\alpha_1, \alpha_2) = \hat{\mu}(t,\alpha_1) - \hat{\mu}(t,\alpha_2)$.

The proposition below shows that if the group level propensity scores and the individual censoring probabilities are known, then the proposed IPCW estimators are unbiased. A proof of the proposition is given in Appendix A.
\begin{prop*} If  $\Pr (\mathbf{A}_i|\mathbf{L}_i)$ and $S_C(X_{ij}|\mathbf{L}_{i}, \mathbf{A}_i)$ are known for  $j = 1,2,\ldots,n_i$ and $i=1,\ldots,m$, and 
then $E\{\hat{\mu}(t,a,\alpha)\} = \mu(t,a,\alpha)$ and $E\{\hat{\mu}(t,\alpha)\} = \mu(t,\alpha)$. \label{paper_1_prp:1}
\end{prop*}

In observational studies,  the conditional distribution of treatment given covariates is unknown.
Likewise, in both observational studies as well as randomized trials, the conditional distribution of censoring given covariates is typically not known (one exception being studies or trials without drop-out such that the only cause of censoring is  the end of administrative follow-up at some fixed time point).
  Therefore, we consider finite dimensional parametric models to estimate the group propensity scores and conditional probability of censoring; these estimates are then plugged into the IPCW estimators defined above. 
  
  The conditional probability of censoring is estimated using a shared frailty model  \citep{munda2012parfm} where the conditional hazard for $C_{ij}$ is assumed to have the proportional hazards form $g_{ij}(c|\mathbf{L}_{i},\mathbf{A}_i,e_i) = g_0(c;\boldsymbol{\theta_h})e_i\exp{(\tilde{\mathbf{L}}_{ij}^T\boldsymbol{\theta_c})},$ where $g_0$ is the baseline hazard function, $\boldsymbol{\theta_h}$ is the $q'$- dimensional parameter vector of the baseline hazard function, $e_i$ is a random effect with density $f_e(e_i;\theta_r)$,
 $\tilde{\mathbf{L}}_{ij}$ is some user specified function of $\{\mathbf{L}_i, \mathbf{A}_i\}$, 
  and $\boldsymbol{\theta_c}$ is the $q$-dimensional vector of coefficients. 
  The  vector $\tilde{\mathbf{L}}_{ij}$ could include, for example, covariates and treatment for individual $j$ (i.e., $\mathbf{L}_{ij}$ and $A_{ij}$) as well as the proportion of others in the group who receive treatment (i.e., $\sum_{k \neq j} A_{ik}/(n_i-1)$).
  Below the dependence of $g_0$ on $\boldsymbol{\theta_h}$ is suppressed for notational convenience. Let  $\boldsymbol{\gamma} = (\boldsymbol{\theta_c}, \boldsymbol{\theta_h}, \theta_r)$ be the vector of parameters for the frailty model. Maximum likelihood theory can be used to draw inference about $\boldsymbol{\gamma}$. Under assumption III, the contribution of group $i$ to the log-likelihood corresponding to the frailty censoring model is \citep{munda2012parfm}
\begin{equation*}
l(\mathbf{X}_i, \boldsymbol{\Delta}_i,\mathbf{L}_i,\mathbf{A}_i,\boldsymbol{\gamma}) = \sum_{j=1}^{n_i}\Delta_{ij}\left[\log\{g_0(X_{ij})\}+\tilde{\mathbf{L}}_{ij}^T\boldsymbol{\theta_c}\right] +(-1)^{d_i}\mathcal{L}^{(d_i)} \sum_{j=1}^{n_i}G_0(X_{ij})\exp{(\tilde{\mathbf{L}}_{ij}^T\boldsymbol{\theta_c})},
\end{equation*}
where $d_i = \sum_{j=1}^{n_i}(1 - \Delta_{ij})$ is the number of censored observations in group $i$, $G_0(\omega) = \int_0^{\omega} g_0(\kappa)d\kappa$, and $\mathcal{L}^{(s)} =\int_0^{\infty}
\exp{(-e_is)}f_e(e_i;\theta_r) de_i.$ Therefore, the maximum likelihood estimator of $\boldsymbol{\gamma}$ solves the following estimating equations
$$\sum_i \psi_{ck}(\mathbf{X}_i, \boldsymbol{\Delta}_i,\mathbf{L}_i,
\mathbf{A}_i,
\boldsymbol{\gamma}) = 0 \text{ for } k = 1,...,q+q'+1,$$
where $\psi_{ck} = \psi_{ck}(\mathbf{X}_i, \boldsymbol{\Delta}_i,\mathbf{L}_i,
\mathbf{A}_i,
\boldsymbol{\gamma}) = \partial l(\mathbf{X}_i, \boldsymbol{\Delta}_i,\mathbf{L}_i,
\mathbf{A}_i,
\boldsymbol{\gamma})/\partial\gamma_k$ and $\gamma_{k}$ is the $k$-th element of $\boldsymbol{\gamma}$.
Below, the baseline hazard for the censoring model is assumed to be constant and equal to $\theta_h$, and the frailty term $e_i$ is assumed to follow a Gamma distribution with mean $1$ and variance $\theta_r$, such that  censoring weights for an uncensored individual can be computed via 
\begin{flalign*}
S_C(t |\mathbf{L}_{i},\mathbf{A}_i,\boldsymbol{\gamma}) &= \int \Pr(C_{ij} >t|\tilde{\mathbf{L}}_{ij}, \boldsymbol{\gamma},e_i ) f_e(e_i;\theta_r)de_i\\
& =  \int \exp{\{-\theta_h  t \exp{(\tilde{\mathbf{L}}_{ij} \boldsymbol{\theta_c}) e_i}\}} \frac{e_i^{1/\theta_r - 1}e^{-e_i/\theta_r}}{\theta_r^{1/\theta_r}\Gamma{(1/\theta_r)}}de_i\\
& =  \left\{\frac{1}{\theta_r \theta_h  t\exp{(\tilde{\mathbf{L}}_{ij} \boldsymbol{\theta_c})}+1}\right\}^{1/\theta_r}
\end{flalign*}

 Following TV (2012), a mixed effects model may be assumed for the treatment allocation, i.e., $\Pr(A_{ij} = 1|\mathbf{L}_{ij},b_i) = \mbox{logit}^{-1}(\mathbf{L}_{ij}\boldsymbol{\theta_x} + b_i)$ where $b_i$ is a random effect following density $f_b(b_i;\theta_s)$. (In the application below the mixed effects model has a slightly more complicated form owing to the particulars of the design of the study analyzed.) Let $\boldsymbol{\beta} = (\boldsymbol{\theta_x},\theta_s)$ denote the $(p+1)$ dimensional vector of parameters for the mixed effects model. Again, maximum likelihood theory can be used to draw inference about $\boldsymbol{\beta}$. The contribution of group $i$ to the log-likelihood for the mixed effects model is given by
$$l(\mathbf{A}_i,\mathbf{L}_i,\boldsymbol{\beta}) = \log {\left[\int \prod_{j=1}^{n_i}h_{ij}(b_i,\mathbf{L}_i,\boldsymbol{\theta_x})^{A_{ij}}\{1-h_{ij}(b_i,\mathbf{L}_i,\boldsymbol{\theta_x})\}^{(1-A_{ij})}f_b(b_i;\theta_s) \right]},$$ 
where $h_{ij}(b_i,\mathbf{L}_i,\boldsymbol{\beta}) = \Pr(A_{ij} = 1|\mathbf{L}_{ij},b_i)$. The maximum likelihood estimator of $\boldsymbol{\beta}$ is the solution to the score equations
$$\sum_i \psi_{xk}(\mathbf{A}_i,\mathbf{L}_i,\boldsymbol{\beta}) = 0 \text{ for } k = 1,...,p+1,$$
where $\psi_{xk} = \psi_{xk}(\mathbf{A}_i,\mathbf{L}_i,\boldsymbol{\beta}) = \partial l(\mathbf{A}_i,\mathbf{L}_i,\boldsymbol{\beta})/\partial\beta_{k}$ and $\beta_{k}$ is the $k$-th element of $\boldsymbol{\beta}$.

Inference about the causal effects of interest is then based on solving the vector of estimating equations
\begin{equation}
\sum_i \psi(\mathbf{O}_i,\boldsymbol{\theta})=0,
\end{equation}
\sloppy where $\boldsymbol{\theta} = (\boldsymbol{\gamma},\boldsymbol{\beta},\theta)$, $\psi(\mathbf{O}_i,\boldsymbol{\theta})= \left( \boldsymbol{\psi_{c}},\boldsymbol{\psi_{x}}, \psi_{a\alpha} \right)^T$, $\boldsymbol{\psi_{c}} = \left(\psi_{c1}, \psi_{c2},..., \psi_{cq+q'+1}\right)^T,$ $\boldsymbol{\psi_{x}} = \left(\psi_{x1}, \psi_{x2},...,\psi_{xp+1}\right)^T,$ 
$$\psi_{a\alpha} = \psi_{a\alpha}(\mathbf{O}_i,\boldsymbol{\theta}) = \frac{g^*(\mathbf{O}_i,a,\alpha,\boldsymbol{\gamma})}{\Pr(\mathbf{A}_i|\mathbf{L}_i,\boldsymbol{\beta})}-\theta,$$
and 
$$
g^*(\mathbf{O}_i,a,\alpha,\boldsymbol{\gamma}) = n_i^{-1}\sum_{j=1}^{n}\frac{\pi (\mathbf{A}_{i,-j};\alpha)I(A_{ij}=a)I(X_{ij}\leq t)}{S_C(X_{ij}|\mathbf{L}_{i},\mathbf{A}_i,\boldsymbol{\gamma})}.
$$ 
Let $\boldsymbol{\hat{\theta}} = (\boldsymbol{\hat{\gamma}},\boldsymbol{\hat{\beta}},\hat{\mu}(t,a,\alpha))$ denote the solution to (1). Denote the true value of $\mathbf{\boldsymbol{\theta}}$ by $\mathbf{\boldsymbol{\theta}_0} = (\boldsymbol{\gamma}_0,\boldsymbol{\beta}_0,\mu(t,a,\alpha))$ and note that
$$\int \psi_{a\alpha}(\mathbf{o},\boldsymbol{\gamma}_0,\boldsymbol{\beta}_0,\mu(t,a,\alpha))dF_\mathbf{O}(\mathbf{o}) = E\left\{\frac{g^*(\mathbf{O}_i,a,\alpha,\boldsymbol{\gamma}_0)}{\Pr(\mathbf{A}_i|\mathbf{L}_i,\boldsymbol{\beta}_0)}-\mu(t,a,\alpha)\right\} = 0,$$
\sloppy where $F_\mathbf{O}$ denotes the joint distribution of the complete observed random variable $\mathbf{O}$ and the last equality follow from the Proposition above. Therefore, assuming the parametric models above are correctly specified, it follows that
$\int \psi(\mathbf{o},\mathbf{\boldsymbol{\theta}}_0)dF_O(\mathbf{o})=0$. By M-estimation theory \citep{stefanski2002calculus}, $\hat{\boldsymbol{\theta}} \overset{p}{\to} \boldsymbol{\theta_0}$ and $\sqrt{m}(\hat{\boldsymbol{\theta}}- \boldsymbol{\theta_0})$ converges in distribution to a Normal distribution with mean $0$ and covariance matrix $\boldsymbol{\Sigma}$ equal to
$U(\boldsymbol{\theta}_0)^{-1}V(\boldsymbol{\theta}_0)\{U(\boldsymbol{\theta}_0)^{-1}\}^T$ where $U(\boldsymbol{\theta}_0) = E\{-\dot{\psi}(\mathbf{O}_i,\mathbf{\boldsymbol{\theta}_0})\}$, $V(\boldsymbol{\theta}_0) = E\{\psi(\mathbf{O}_i,\mathbf{\boldsymbol{\theta}_0})\psi(\mathbf{O}_i,\mathbf{\boldsymbol{\theta}_0})^T\}$, and $\dot{\psi}(\mathbf{O}_i,\boldsymbol{\theta}) = \partial \psi(\mathbf{O}_i,\mathbf{\boldsymbol{\theta}})/\partial\mathbf{\boldsymbol{\theta}}^T$. 
Consistency and asymptotic normality of the direct, indirect and total effect estimators follows from the delta method. Similar techniques can be used to show that $\hat{\mu}(t,\alpha)$ and the overall effect estimator are also consistent and asymptotically Normal. The asymptotic variance $\boldsymbol{\Sigma}$ can be consistently estimated by $\hat{\boldsymbol{\Sigma}} = \hat{U}(\hat{\boldsymbol{\theta}})^{-1}\hat{V}(\hat{\boldsymbol{\theta}})\{\hat{U}(\hat{\boldsymbol{\theta}})^{-1}\}^T$ 
where $\hat{U}(\hat{\boldsymbol{\theta}}) = m^{-1}\sum_{i = 1}^{m}\{-\dot{\psi}(\mathbf{O}_i,\hat{\boldsymbol{\theta}})\}$ and $\hat{V}(\hat{\boldsymbol{\theta}}) = m^{-1}\sum_{i=1}^{m}\{\psi(\mathbf{O}_i,\hat{\boldsymbol{\theta}})\psi(\mathbf{O}_i,\hat{\boldsymbol{\theta}})^T\}.$ The empirical sandwich variance estimator $\hat{\boldsymbol{\Sigma}}$ can be computed using the R package \texttt{geex} \citep{saul2019} and can be used to construct Wald type confidence intervals (CIs).

\section{Simulation Study}

A simulation study was conducted to assess the finite sample bias of the IPCW estimator and coverage of the corresponding Wald confidence intervals. The data generating model used in the simulation study was motivated by aspects of the cholera vaccine study analysis presented in the next section. Following \cite{perez2014assessing}, data were simulated according to the following steps.
\begin{enumerate}[i)]
\item  First, two baseline covariates $L_{1ij}$ and $L_{2ij}$ were randomly generated. In the application presented in Section 4, conditional independence (assumption I) is assumed given an individual's age (in decades) and the distance of their residence to the nearest river. Motivated by this example, $L_{1ij}$ (age) and $L_{2ij}$ (distance to river) were randomly generated as follows. First, $V_{ij}$ was randomly generated from an Exponential distribution with mean $20$,
$r_{1i}$ from Normal$(0,0.1)$, and $r_{2ij}$ from Normal$(0,0.1)$. 
Then $L_{1ij}$ was set to $\min(V_{ij}+r_{1i}+r_{2ij},100)/10$ and $L_{2ij}$ randomly generated such that $\log L_{2ij}\sim \mbox{Normal}(r_{1i}+r_{2ij},0.75).$
\item The random effects for the treatment model $b_i$ were randomly sampled from a Normal distribution with mean $0$ and variance $0.0859$. 
\item The treatment indicators $A_{ij}$ were randomly sampled from a Bernoulli distribution with mean $p_{ij} = \mbox{expit}(0.2727-0.0387 L_{1ij} + 0.2179 L_{2ij} + b_i)$.
\item The potential times to event $T_{ij}(\mathbf{a}_{i})$ were randomly sampled from an Exponential distribution with mean $\mu_{ij} = 200+100a_{ij}-0.98L_{1ij} -  0.145 L_{2ij} + 50\sum_{k\neq j}a_{ik}/n_i$.
\item  The random effects for the censoring model $e_i$ were randomly generated from a Gamma distribution with mean $1$ and variance $\theta = 1.25.$
\item Censoring times $C_{ij}$ were randomly sampled from an Exponential distribution with mean $1/\lambda_0$ where $\lambda_0 = 0.015\exp{(0.002L_{1ij}+ 0.015L_{2ij})}e_i$.
\item Individual censoring indicators were determined, i.e., $\Delta_{ij} = 0$ if $C_{ij}<T_{ij}(\mathbf{A}_i)$.

\end{enumerate}

Steps i through vii were used to stochastically generate 1000 data sets, with each data set containing 500 groups with 10 individuals per group.  For each simulated data set, the IPCW estimator of $\mu(100,a,\alpha)$ was evaluated for $a=0,1$ and $\alpha = 0.1, 0.2, \ldots, 0.9$. Estimated standard errors based on the empirical sandwich variance estimator and Wald 95\% confidence intervals
were also calculated for each simulated data set. Empirical standard errors were calculated by taking the standard deviation of the point estimates from all simulations.
\par The true value of the estimand was obtained by simulating counterfactual outcomes for $m = 10^6$ groups of individuals. Note that, according to the model used to generate the data, potential survival times depend only on $\sum_{k\neq j} a_{ik}$. So, $\mu(t,a,\alpha)$ was approximated by \citep{perez2014assessing} 
$$m^{-1}\sum_{i=1}^{m}n_i^{-1}\sum_{j=1}^{n_i}\sum_{k = 0}^{n_i - 1} {{n_i-1}\choose{k}} I \{ T_{ij}(a,k)\leq t\} \alpha^k(1 - \alpha)^{n_i - k - 1}.$$ 
The true value of $\mu(t,\alpha)$ was determined in a similar fashion.
\par Results from the simulation study are presented in Table $1$. Bias of the IPCW estimator was negligible for all values of $a$ and $\alpha$. Likewise, the average estimated standard error was close to the empirical standard error. Coverage of the $95\%$ Wald CIs was approximately equal to the nominal level.

\begin{table}
\resizebox{\columnwidth}{!}{
\begin{tabular}{ccccccccccccc}
\cline{1-6} \cline{8-13}
$\alpha$ & $\mu(100,0,\alpha)$ & Bias & ESE & ASE  & EC & & $\alpha$ & $\mu(100,1,\alpha)$ & Bias & ESE & ASE & EC \\ 
\cline{1-6} \cline{8-13}
0.1 & 0.39 & 0.02 & 0.07 & 0.07 & 94$\%$ & & 0.1 &  0.28 & 0.01 & 0.08 & 0.08 & 92$\%$\\ 
0.2 & 0.38 & 0.01 & 0.04 & 0.04 & 96$\%$ & & 0.2 & 0.27 & 0.01 & 0.04 & 0.04 & 95$\%$ \\ 
0.3 & 0.38 & 0.00 & 0.03 & 0.03 & 96$\%$ & & 0.3 & 0.27 & -0.00 & 0.03 & 0.03 & 95$\%$ \\ 
0.4 & 0.37 & -0.00 & 0.03 & 0.02 & 95$\%$ & & 0.4 & 0.27 & -0.01 & 0.02 & 0.02 & 94$\%$  \\ 
0.5 & 0.36 & -0.00 & 0.03 & 0.02 & 94$\%$ & & 0.5 & 0.26 & -0.00 & 0.02 & 0.02 & 93$\%$ \\ 
0.6 & 0.36 & -0.01 & 0.03 & 0.02 & 94$\%$ & & 0.6 &  0.26 & -0.00 & 0.02 & 0.02 & 93$\%$ \\ 
0.7 & 0.35 & -0.00 & 0.03 & 0.02 & 94$\%$ & & 0.7 & 0.26 & -0.01 & 0.02 & 0.01 & 94$\%$ \\ 
0.8 & 0.35 & -0.01 & 0.03 & 0.03\ & 94$\%$ & & 0.8 & 0.25 & -0.00 & 0.02 & 0.02 & 93$\%$ \\ 
0.9 & 0.34 & -0.00 & 0.05 & 0.05 & 92$\%$ & & 0.9 & 0.25 & 0.01 & 0.02 & 0.02 & 95$\%$ \\ 
\cline{1-6} \cline{8-13}
\end{tabular} 
}\caption{Results from simulation study described in Section $3$. $\alpha$ denote the allocation probabilities, $\mu(100, a, \alpha)$ is the true value of the target parameter for $a = 0,1$; Bias is the average of $\mu(100, a, \alpha) - \hat{\mu}(100, a, \alpha)$ for $a = 0,1$; ESE is the empirical standard error; ASE is the average of the sandwich variance based standard error estimates; and EC denotes the empirical coverage of the $95\%$ Wald confidence intervals.}\label{paper_1_table_1}
\end{table}

\par Additional simulation studies were conducted to assess the performance of the proposed methods for different values of $m$, the total number of groups, ranging from 10 to 500. The number of individuals per group was 10, as in the previous simulations. For each $m \in \{10,50,100,200,300,400,500\}$, $1000$ data sets were simulated according to steps i through vii above. Results are depicted in Figure 1. Bias of the IPCW estimator was  small and coverage of the Wald CIs was close to the nominal level provided $m$ was at least 50. Additional details of these simulation results are provided in Appendix Tables 1--5.
 In cluster randomized trials with small numbers of clusters, Wald-type CIs are often constructed using a $t$ distribution with $m-r$ degrees of freedom, where $r$ is the number of parameters being estimated, rather than a Normal distribution;
empirical coverage of  Wald CIs using the $t$ distribution is also shown in Appendix Tables 1--5 and was similar to coverage based on the Normal distribution.

\begin{figure}
\caption{Absolute bias (left) and $95\%$ confidence interval coverage (right) for different numbers of groups for $\alpha = 0.5$. The dotted line in the right plot corresponds to $95\%$ coverage.}\label{paper_com_sim}
\centering
\includegraphics[width=\textwidth]{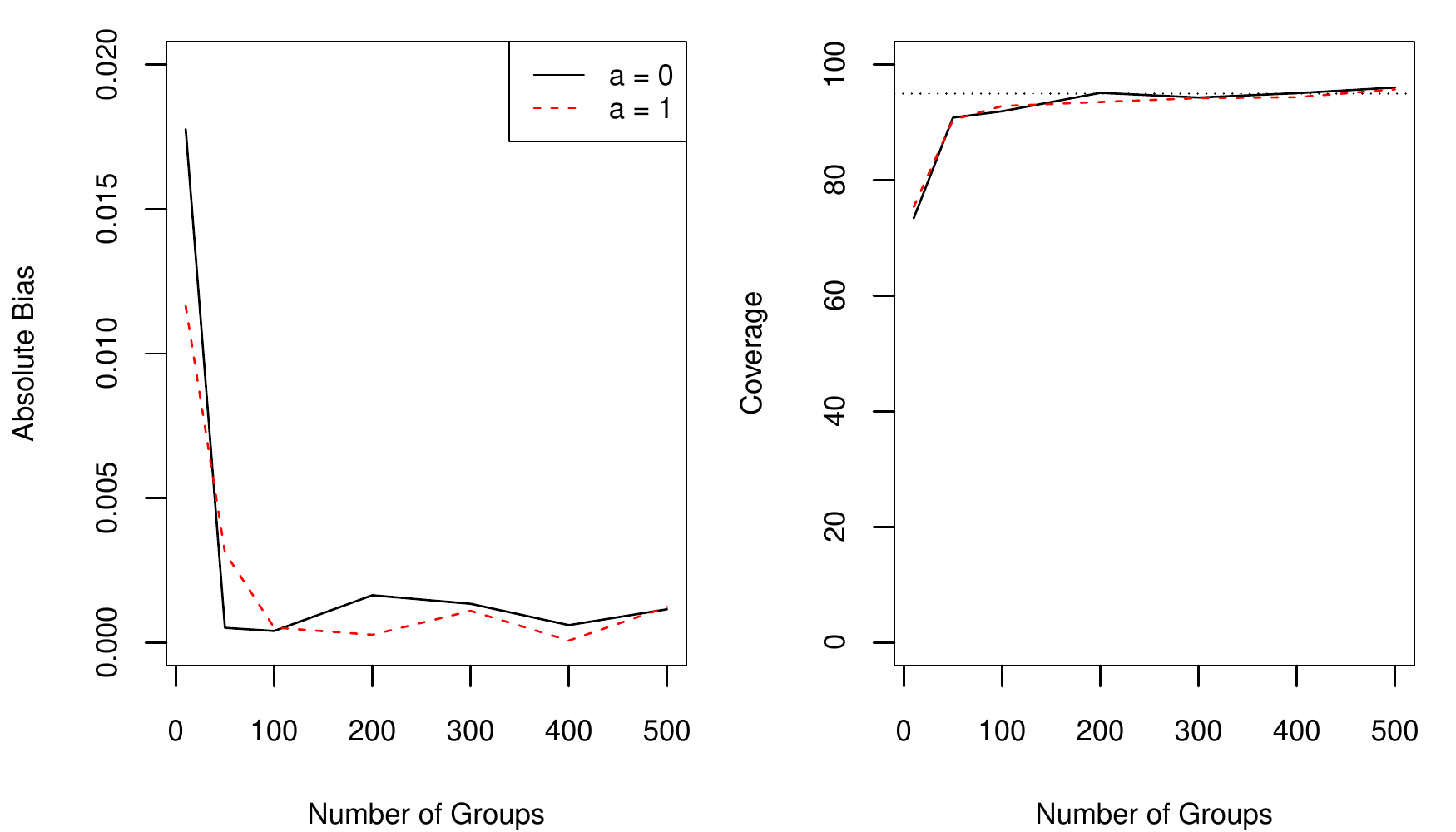}
\end{figure}

\section{Application}
\subsection{Cholera Vaccine Study and Analysis} 
In this section, the methods described in Section 2 are used to analyze a cholera vaccine study in Matlab, Bangladesh \citep{ali2005herd}. Eligible study participants were children 2--15 years of age and women greater than 15 years old. All $121,975$ eligible individuals in the population were randomized to one of three vaccination groups: B subunit-killed whole-cell oral cholera vaccine, killed whole-cell-only cholera vaccine, and {\it E. coli} K12 placebo. As in Perez-Heydrich et al.\ (2014), no distinction is made between the two vaccines in the analysis presented here. Individuals were considered to have participated in the randomized trial component of the study if they received two or more doses of vaccine or placebo. The primary endpoint of the trial was incident cholera. Three health centers in the Matlab area served as surveillance centers and collected endpoint data on all individuals, regardless of whether they participated in the randomized trial. The analysis presented here includes data from all individuals, i.e., trial participants as well as those who chose not to participate. Thus an approach which accounts for possible confounding, such as the IPW method described in Section 2, should be utilized to assess the effects of vaccination.

\par Previous analyses of this study suggest the presence of interference \citep{ali2005herd, perez2014assessing}. 
Interference is plausible in the setting because the vaccine may (i) prevent an individual from contracting cholera or (ii) decrease the infectiousness or contagiousness of an individual who does contract cholera; for either (i) or (ii),  the vaccine would make it less likely that such an individual would subsequently infect other individuals. 
However, these previous analyses did not formally account for censoring. Here individuals are considered right censored if they were not diagnosed with cholera during the study. Individuals who emigrated from the study location or died during the follow-up period prior to cholera infection were right censored at the time of emigration or death. Individuals who did not emigrate or die and who did not develop cholera during the study were right censored at the end of the study period.

\par Related individuals in Matlab live in clustered sets of houses called baris. There were a total of 6,415 baris at the time of the vaccine trial. \cite{perez2014assessing} used a clustering algorithm to form groups (neighborhoods) based on the spatial location of the baris, with the number of groups pre-specified to be 700. The analysis here is based on the same groups as in Perez-Heydrich et al. and assumes that there is no interference between individuals in different groups, i.e, the vaccination of an individual in one group has no effect on whether an individual in another group contracts cholera. When fitting the propensity model $\Pr (\mathbf{A}_i|\mathbf{L}_i,\boldsymbol{\beta})$ described below, the largest 15 groups had estimated group propensity scores that were effectively equal to zero and therefore these groups were omitted.

\par Individuals participating in the vaccine trial were not all vaccinated on the same calendar day, such that the level of vaccine coverage within a group varied over a relatively brief period of calendar time at the study onset. For simplicity and because the methods developed above do not accommodate time varying treatment, the start of follow-up for all individuals in a particular group was set to the latest date of second vaccination among all individuals in that group. Observations were excluded if individuals contracted cholera (60), died (346), or emigrated  (3671) prior to the start of follow-up for their group. In total, 94,234 individuals were included in the analysis. Among these individuals, 55,413 were unvaccinated, either because they received placebo or they did not participate, and 38,821
were vaccinated with one of the two vaccines. During follow-up, there were 280 incident cases of cholera among the unvaccinated individuals and 74 cholera cases among the vaccinated individuals.

\par As in Perez-Heydrich et al., the group propensity score was modeled using a mixed effects model. The particular form of the model derives from the fact that in order for an individual  to have received a vaccine, they must have (i) chosen to participate in the trial, and (ii) been randomized to receive one of the two vaccines. To account for (i), a logistic regression model for participation was assumed. As in Perez-Heydrich et al., covariates in the participation component of the model were age, squared age, distance to nearest river, and squared distance to nearest river. Accommodating (ii) in the propensity model is straightforward because, due to randomization, individuals who elected to participate in the trial were known to receive one of the two vaccines with probability 2/3. Combining these two aspects of the model, the propensity score for group $i$ was estimated by
$$\Pr (\mathbf{A}_i|\mathbf{L}_i,\boldsymbol{\hat{\beta}}) = \int \prod_{j=1}^{n_i}\{ (2/3)h_{ij}(b_i,\mathbf{L}_{ij},\boldsymbol{\hat{\theta}_x)}\}^{A_{ij}}\{ 1-(2/3)h_{ij}(b_i,\mathbf{L}_{ij},\boldsymbol{\hat{\theta}_x})\}^{(1-A_{ij})}f_b(b_i;\hat{\theta}_s),$$
where $h_{ij}(b_i,\mathbf{L}_i,\boldsymbol{\theta}_x) = \Pr(B_{ij}=1|b_i,\mathbf{L}_{ij},\boldsymbol{\theta}_x) = \mbox{expit}(\mathbf{L}_{ij}\boldsymbol{\theta_x} + b_i)$, $B_{ij}$ is the indicator of participation, i.e., $B_{ij} = 1$ if individual $j$ in group $i$ participated in the randomized trial and $B_{ij} = 0$ otherwise, and $(\boldsymbol{\hat{\theta}_x}, \hat \theta_s)$ is the maximum likelihood estimate of $(\boldsymbol{\theta_x}, \theta_s)$. Censoring was modeled using the Gamma frailty model described above, and only included age as a covariate as no other variables were associated with censoring. Over $70\%$ of individuals belonged to groups where the vaccine coverage was between $0.3$ and $0.6$. Therefore, the analysis was conducted for allocation strategies ranging from $0.3$ to $0.6$.

\subsection{Results}
\begin{figure}
\caption{Estimated cumulative probability of cholera over time when vaccinated or unvaccinated for $\alpha = 0.3$ (left), $\alpha = 0.45$ (center) and $\alpha = 0.6$ (right)}\label{paper_1_fig_1}
\centering\resizebox{\columnwidth}{!}{
\includegraphics[width=\textwidth]{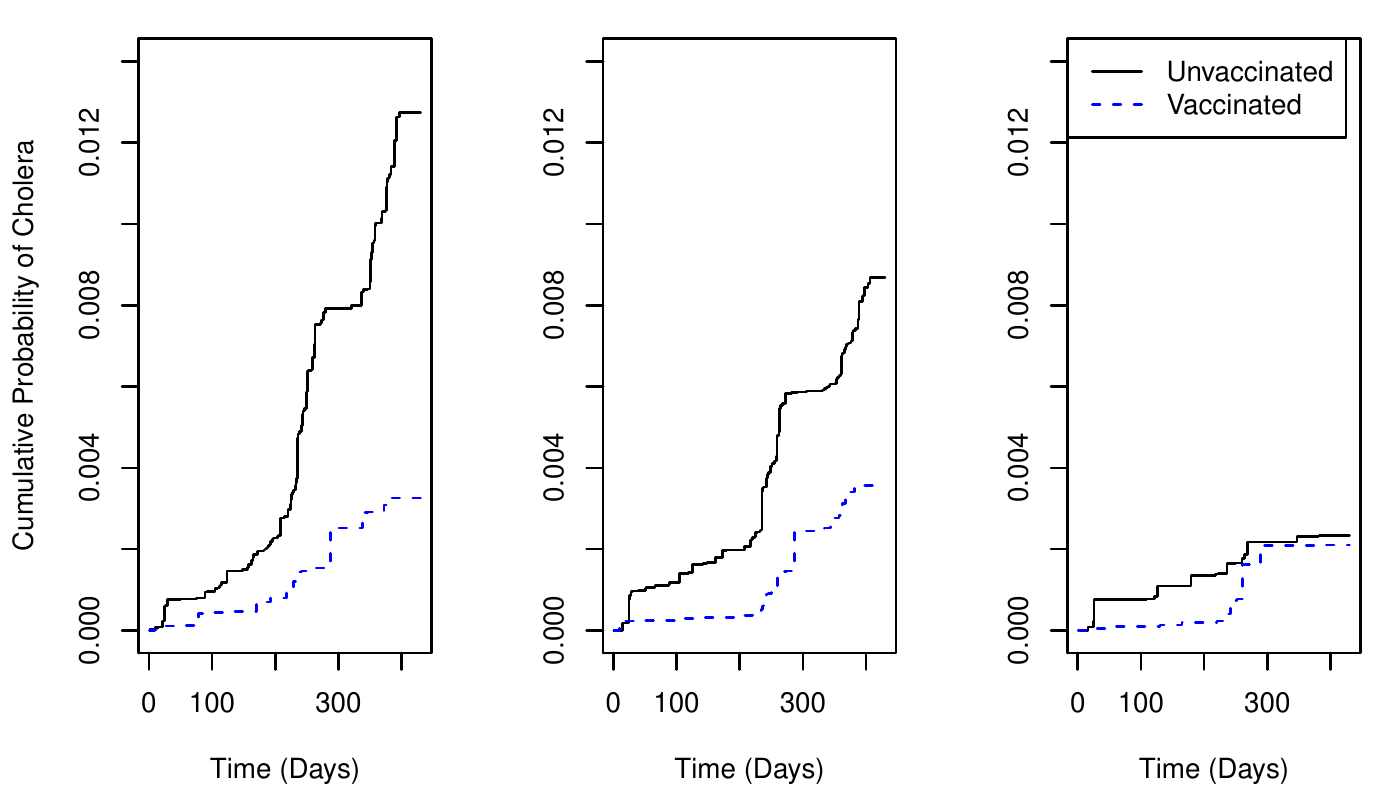}
}
\end{figure}
\par
Figure \ref{paper_1_fig_1} shows the IPCW estimates of the cumulative probability of cholera over time for allocation
strategies 0.3, 0.45, and 0.6, both when an individual receives a vaccine and when an individual is unvaccinated. 
Figure 1 shows that 
the estimated risk of cholera when an individual is unvaccinated decreases dramatically as $\alpha$ increases, suggesting the presence of interference. This decrease is more modest when an individual is vaccinated, indicating a stronger indirect effect when unvaccinated. At all time points the estimated risk of cholera is higher when an individual is unvaccinated, suggesting a beneficial, direct effect of vaccination, especially at lower coverage levels. For $\alpha=0.3$ and $\alpha=0.45$, the estimated risk when unvaccinated increases suddenly between 200 and 300 days, and then
again between 300 and 400 days. These results might be attributable to the known bimodal seasonality of cholera in Bangladesh  \citep{longini2002epidemic}. Note that, because the study start date varied across groups, the time scale in this analysis does not exactly coincide with calendar time. Nonetheless, $95\%$ of individuals had a start date within a two calendar month range, such that there is a strong correlation between the analysis time scale  and calendar time, and thus cholera seasonality may explain these periods of marked increase in risk.

\begin{figure}[h!]
\caption{Direct, indirect, total and overall effect estimates ($\times 1000$) for different allocation strategies at time $t = 1$ year. Indirect, total, and overall effects are with respect to $\alpha_2 = 0.4$. The shaded regions denote pointwise 95\% confidence intervals of the estimates.}\label{paper_1_fig_2}
\centering
\resizebox{\columnwidth}{!}{
\includegraphics[width=1.4\textwidth]{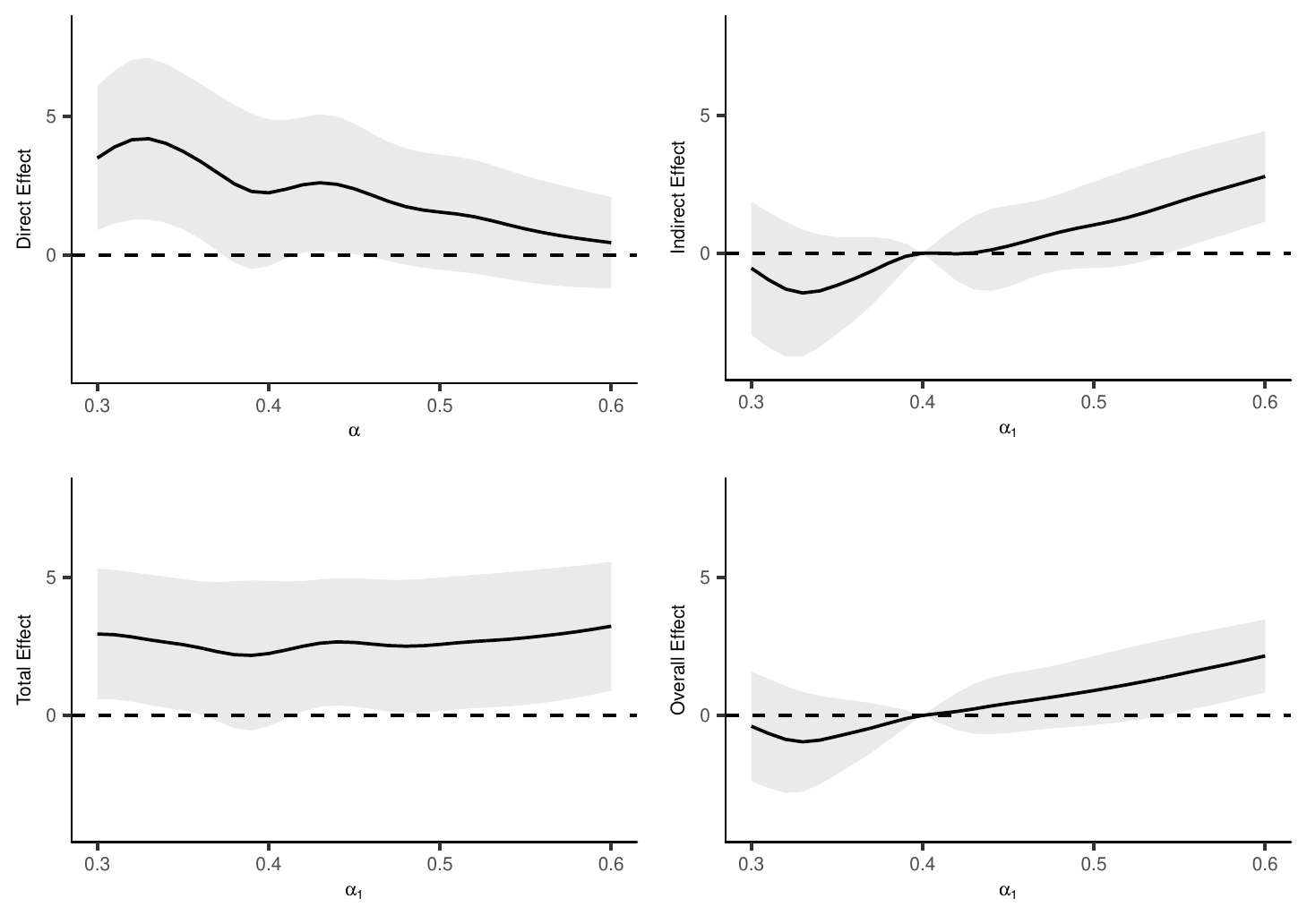}
}
\end{figure}

\par
Direct, indirect, total and overall effect estimates and $95\%$ CIs ($\times1000$) for different allocation strategies at time $t = 1$ year are shown in Figure \ref{paper_1_fig_2}. The direct effect estimates generally decrease as $\alpha$ increases. For example, the direct effect estimate for $\alpha = 0.35$ is $3.6$ $(95\% $ CI $1.1, 6.2)$ whereas for $\alpha = 0.5$ the direct effect estimate is $1.5$ $(95\% $ CI $-0.5, 3.5)$. The indirect, total, and overall effect estimates in Figure \ref{paper_1_fig_2} compare the risk of cholera over a range of allocation probabilities $\alpha_1 \in [0.3, 0.6]$ versus $\alpha_2=0.4$. Here the indirect effect contrasts risk of cholera infection when individuals are unvaccinated. For larger values of $\alpha_1$ the $95\%$ CIs for these effects exclude the null value of zero. For example, for $\alpha_1=0.6$ the indirect effect estimate is 2.8 ($95\%$ CI 1.1, 4.5), providing statistical evidence of the presence of interference. These results indicate that when individuals are unvaccinated, the risk of cholera infection is significantly reduced by increasing the level of vaccine coverage in their neighborhood. The total effect estimates quantify the combined direct and indirect effects of the vaccine. The overall effect estimates may be of greatest interest from a public health or policy perspective. For $\alpha_1=0.6$, the overall effect estimate is 2.2 ($95\%$ CI 0.9, 3.4); in words, 2.2 fewer cases of cholera per 1000 individuals per year are expected if $60\%$ of individuals are vaccinated compared to if only $40\%$ of individuals receive vaccine. 
\par In previous analyses of these data, Perez-Heydrich et al.\ also estimated the direct, indirect, total and overall effects using a binary outcome indicating whether an individual was infected with cholera during the first year of follow-up. The IPCW estimates for $t=1$ are similar to these previous results, e.g., Perez-Heydrich et al.\ estimated the direct effect for $\alpha=0.32$ to be 5.3 ($95\%$ CI 2.5, 8.1) whereas the IPCW estimate of this effect at $t=1$ is 4.0 ($95\%$ CI 1.6, 6.5).  However, the  Perez-Heydrich et al.\ estimates may be biased because they did not account for right censoring.
That said, the IPCW results should still be viewed cautiously and only have valid causal interpretation if Assumptions I - IV hold and the treatment and censoring models are correctly specified.

\section{Discussion}

In this paper, the TV IPW estimator for partial interference was extended to allow for right censored outcomes. The proposed estimator was obtained by weighting the original TV estimator by censoring weights calculated from a parametric frailty model of the censoring times. 
The estimator was shown to be consistent and asymptotically Normal (under identifiability Assumptions I - IV), and a consistent estimator of the asymptotic variance was proposed. A simulation study demonstrated that the proposed methods performed well in finite samples provided the number of groups is sufficiently large. Analysis of a cholera vaccine study using the proposed methods suggests vaccination had both a direct and indirect effect against cholera infection. These results are in accordance with findings by \cite{ali2005herd} and \cite{perez2014assessing}, but are likely more accurate since these previous analyses did not formally account for right censoring.

There are several areas of possible future research related to the methods developed here.
For example, further research could entail developing estimators which perform well in settings where the number of groups ($m$) is small.
The IPCW estimator also presents computational challenges when group sizes ($n_i$'s) are large because the corresponding estimated group propensity scores can be approximately zero; other approaches are needed to better accommodate larger groups. 
Validity of the IPCW estimator requires correct specification of parametric models, such that it is important to assess model fit in application of this method.  Further research could entail extensions which utilize  semi-parametric data-adaptive methods instead.
 Extensions of the IPCW estimator could also be considered for the setting where there is general interference, i.e., where interference is not restricted to individuals within the same group. In this paper only Horwitz-Thompson type IPCW estimators were considered; further research could entail developing stabilized or Hajek type  \citep{liu2016inverse} IPCW estimators for right censored data which may be more stable and less variable. 

Methods could also be developed allowing for the proportion of treated individuals to vary over time, which perhaps could entail adapting existing methods which  accommodate time varying exposures and confounding but assume no interference between individuals. The approach considered in this paper could also be extended to allow for the possibility of ``covariate interference,'' i.e., the covariate values of one individual could  affect the outcome of another individual \citep{balzer2019,prague2016}.
  Finally, following 
\cite{tchetgen2010causal}
and
\cite{perez2014assessing}, here we consider causal estimands corresponding to counterfactual scenarios where individuals independently select treatment with probability $\alpha$; in future research alternative estimands, such as in \cite{barkley2017}, could be considered which may be more relevant to public health officials determining vaccine policy.

\section*{Acknowledgment}
The authors thank  Brian Barkley, Bradley Saul, Shaina Mitchell and Kayla Kilpatrick for their useful comments and suggestions. This work was supported
by NIH grant R01 AI085073. The content is solely the responsibility of
the authors and does not necessarily represent the official views of the NIH.

\bibliographystyle{apa}
\bibliography{bib}

\section*{Appendix A. Proposition Proof}

From the definition of the IPCW estimator, 
\begin{equation}
E\{\hat{F}_i(t,a,\alpha)\} = E\left\{\sum_{j=1}^{n_i} \frac{\pi(\mathbf{A}_{i,-j};\alpha) I(A_{ij}=a)I(\Delta_{ij}=1)I(X_{ij}\leq t)}{S_C(X_{ij}|\mathbf{L}_{i}, \mathbf{A}_i)\Pr(\mathbf{A}_i|\mathbf{L}_i)n_i}\right\}
\end{equation}
Noting $\Delta_{ij}=1$ if and only if $C_{ij}>T_{ij}(\mathbf{A}_i)$, 
by the law of total expectation and causal consistency the right side of $(1)$ can be expressed as 
$$
E_{\mathbf{L}_i,\mathbf{A}_i,T_{ij}(\mathbf{A}_i)}E_{C_{ij}|\mathbf{L}_i,\mathbf{A}_i,T_{ij}(\mathbf{A}_i)}\left[\sum_{j=1}^{n_i} \frac{\pi(\mathbf{A}_{i,-j};\alpha) I(A_{ij}=a)I\{C_{ij}>T_{ij}(\mathbf{A})\}I\{T_{ij}(\mathbf{A}_i)\leq t\}}{S_C\{T_{ij}(\mathbf{A})|\mathbf{L}_{i},\mathbf{A}_i\}\Pr(\mathbf{A}_i|\mathbf{L}_i)n_i}\right]
$$
Moving the inner expectation inside the summation and taking out terms that are constant with respect to that expectation, 
it follows  that 
$E\{\hat{F}_i(t,a,\alpha)\}$ equals\\
$$
E_{\mathbf{L}_i,\mathbf{A}_i,T_{ij}(\mathbf{A}_i)}\left[\sum_{j=1}^{n_i} \frac{\pi(\mathbf{A}_{i,-j};\alpha) I(A_{ij}=a)E\{I(C_{ij}>T_{ij}(\mathbf{A}_i))|\mathbf{L}_i,\mathbf{A}_i,T_{ij}(\mathbf{A}_i)\}I\{T_{ij}(\mathbf{A}_i)\leq t\}}{S_C\{T_{ij}(\mathbf{A}_i)|\mathbf{L}_{i},\mathbf{A}_i\}\Pr(\mathbf{A}_i|\mathbf{L}_i)n_i}\right]
$$
Note by Assumption III that for any $t$ $$S_c(t|\mathbf{L}_i,\mathbf{A}_i) =   \Pr\{C_{ij}>t|\mathbf{L}_i,\mathbf{A}_i,T_{ij}(\mathbf{A}_i)=t\}
=
E\{I(C_{ij}>t)|\mathbf{L}_i,\mathbf{A}_i,T_{ij}(\mathbf{A}_i)=t
\}
$$
Therefore
\begin{equation*}
E\{\hat{F}_i(t,a,\alpha)\} = E_{\mathbf{A}_i,\mathbf{L}_i,T_{ij}\mathbf{(A}_i)}\left[\frac{\sum_{j=1}^{n_i}\pi(\mathbf{A}_{i,-j};\alpha) I(A_{ij}=a)I\{T_{ij}(\mathbf{A}_i)\leq t\}}{\Pr(\mathbf{A}_i|\mathbf{L}_i)n_i}\right].
\end{equation*}
The remainder of the proof follows from the proof of Theorem 6 of Tchetgen Tchetgen and VanderWeele (2012).

\newpage

\section*{Appendix Tables}

\vspace{1in}

\begin{center}
{\bf Appendix Table 1}\\
{\it Simulation results for $m=50$ and $n_i=10$ for all $i$, where:
$\alpha$ denotes the allocation probabilities; 
 $\mu_a$ denotes denotes the value of the target parameters $\mu(100,a,\alpha)$ for $a=0,1$;
Bias is the average of $\mu(100, a, \alpha) - \hat{\mu}(100, a, \alpha)$ for $a = 0,1$; 
ESE is the empirical standard error; ASE is the average of the sandwich variance based standard error estimates; 
EC denotes the empirical coverage of the $95\%$ Wald confidence intervals based on the Normal distribution;
and EC$_t$ denotes empirical coverage of $t$ distribution-based Wald CIs.}\\
\vspace{.1in}
\begin{tabular}{ccccccccccccccc}
\cline{1-7} \cline{9-15}
$\alpha$ & $\mu_0$ & Bias & ESE & ASE  & EC &EC$_t$ & & $\alpha$ & $\mu_1$ & Bias & ESE & ASE & EC &EC$_t$ \\ 
\cline{1-7} \cline{9-15}
0.1 & 0.39 & -0.01 & 0.25 & 0.17 & 76\% & 77\% && 0.1 & 0.28 & -0.01 & 0.28 & 0.17 & 64\% & 64\% \\
0.2 & 0.38 & -0.01 & 0.13 & 0.11 & 86\% & 87\% && 0.2 & 0.27 & -0.00 & 0.15 & 0.12 & 79\% & 80\% \\
0.3 & 0.37 & -0.00 & 0.10 & 0.09 & 88\% & 89\% && 0.3 & 0.27 & 0.00  & 0.10 & 0.08 & 87\% & 87\% \\
0.4 & 0.37 & 0.00  & 0.08 & 0.07 & 89\% & 90\% && 0.4 & 0.27 & 0.00  & 0.07 & 0.06 & 90\% & 90\% \\
0.5 & 0.36 & 0.01  & 0.07 & 0.07 & 90\% & 91\% && 0.5 & 0.26 & 0.00  & 0.06 & 0.05 & 90\% & 91\% \\
0.6 & 0.36 & 0.01  & 0.07 & 0.06 & 90\% & 91\% && 0.6 & 0.26 & 0.00  & 0.05 & 0.05 & 90\% & 91\% \\
0.7 & 0.35 & 0.01  & 0.07 & 0.07 & 88\% & 89\% && 0.7 & 0.26 & 0.00  & 0.05 & 0.04 & 92\% & 93\% \\
0.8 & 0.35 & 0.01  & 0.10 & 0.09 & 88\% & 89\% && 0.8 & 0.25 & 0.00  & 0.05 & 0.05 & 92\% & 92\% \\
0.9 & 0.34 & 0.00  & 0.16 & 0.13 & 84\% & 84\% && 0.9 & 0.25 & -0.00 & 0.07 & 0.06 & 90\% & 90\%\\
\cline{1-7} \cline{9-15}
\end{tabular}
\end{center}

\newpage{}

\begin{center}
{\bf Appendix Table 2}\\
{\it Simulation results for $m=100$ and $n_i=10$ for all $i$, where:
$\alpha$ denotes the allocation probabilities; 
 $\mu_a$ denotes denotes the value of the target parameters $\mu(100,a,\alpha)$ for $a=0,1$;
Bias is the average of $\mu(100, a, \alpha) - \hat{\mu}(100, a, \alpha)$ for $a = 0,1$; 
ESE is the empirical standard error; ASE is the average of the sandwich variance based standard error estimates; 
EC denotes the empirical coverage of the $95\%$ Wald confidence intervals based on the Normal distribution;
and EC$_t$ denotes empirical coverage of $t$ distribution-based Wald CIs.}\\
\vspace{.1in}
\begin{tabular}{ccccccccccccccc}
\cline{1-7} \cline{9-15}
$\alpha$ & $\mu_0$ & Bias & ESE & ASE  & EC &EC$_t$ & & $\alpha$ & $\mu_1$ & Bias & ESE & ASE & EC &EC$_t$ \\ 
\cline{1-7} \cline{9-15}
0.1 & 0.39 & -0.02 & 0.20 & 0.15 & 83\% & 84\% && 0.1 & 0.28 & -0.01 & 0.20 & 0.14 & 77\% & 78\% \\
0.2 & 0.38 & -0.01 & 0.10 & 0.09 & 91\% & 91\% && 0.2 & 0.27 & -0.01 & 0.10 & 0.09 & 87\% & 87\% \\
0.3 & 0.37 & -0.01 & 0.07 & 0.06 & 91\% & 91\% && 0.3 & 0.27 & -0.00 & 0.07 & 0.06 & 91\% & 91\% \\
0.4 & 0.37 & -0.01 & 0.06 & 0.05 & 93\% & 93\% && 0.4 & 0.27 & 0.00  & 0.05 & 0.05 & 93\% & 94\% \\
0.5 & 0.36 & -0.00 & 0.05 & 0.05 & 93\% & 94\% && 0.5 & 0.26 & 0.00  & 0.04 & 0.04 & 93\% & 93\% \\
0.6 & 0.36 & 0.00  & 0.05 & 0.05 & 92\% & 93\% && 0.6 & 0.26 & 0.00  & 0.04 & 0.03 & 92\% & 93\% \\
0.7 & 0.35 & 0.00  & 0.06 & 0.05 & 92\% & 92\% && 0.7 & 0.26 & 0.00  & 0.04 & 0.03 & 91\% & 91\% \\
0.8 & 0.35 & 0.01  & 0.07 & 0.06 & 90\% & 90\% && 0.8 & 0.25 & -0.00 & 0.04 & 0.03 & 92\% & 93\% \\
0.9 & 0.34 & 0.01  & 0.11 & 0.10 & 88\% & 88\% && 0.9 & 0.25 & -0.01 & 0.05 & 0.05 & 93\% & 93\%\\
\cline{1-7} \cline{9-15}
\end{tabular}
\end{center}

\newpage{}

\begin{center}
{\bf Appendix Table 3}\\
{\it Simulation results for $m=200$ and $n_i=10$ for all $i$, where:
$\alpha$ denotes the allocation probabilities; 
 $\mu_a$ denotes denotes the value of the target parameters $\mu(100,a,\alpha)$ for $a=0,1$;
Bias is the average of $\mu(100, a, \alpha) - \hat{\mu}(100, a, \alpha)$ for $a = 0,1$; 
ESE is the empirical standard error; ASE is the average of the sandwich variance based standard error estimates; 
EC denotes the empirical coverage of the $95\%$ Wald confidence intervals based on the Normal distribution;
and EC$_t$ denotes empirical coverage of $t$ distribution-based Wald CIs.}\\
\vspace{.1in}
\begin{tabular}{ccccccccccccccc}
\cline{1-7} \cline{9-15}
$\alpha$ & $\mu_0$ & Bias & ESE & ASE  & EC &EC$_t$ & & $\alpha$ & $\mu_1$ & Bias & ESE & ASE & EC &EC$_t$ \\ 
\cline{1-7} \cline{9-15}
0.1 & 0.39 & -0.01 & 0.12 & 0.11 & 87\% & 87\% && 0.1 & 0.28 & -0.01 & 0.13 & 0.11 & 86\% & 86\% \\
0.2 & 0.38 & -0.01 & 0.06 & 0.06 & 94\% & 94\% && 0.2 & 0.27 & -0.01 & 0.07 & 0.07 & 91\% & 91\% \\
0.3 & 0.37 & -0.01 & 0.05 & 0.04 & 95\% & 95\% && 0.3 & 0.27 & -0.00 & 0.05 & 0.04 & 93\% & 93\% \\
0.4 & 0.37 & -0.00 & 0.04 & 0.04 & 93\% & 94\% && 0.4 & 0.27 & 0.00  & 0.03 & 0.03 & 94\% & 94\% \\
0.5 & 0.36 & 0.00  & 0.04 & 0.03 & 93\% & 93\% && 0.5 & 0.26 & 0.00  & 0.03 & 0.03 & 94\% & 94\% \\
0.6 & 0.36 & 0.00  & 0.04 & 0.03 & 94\% & 94\% && 0.6 & 0.26 & 0.00  & 0.02 & 0.02 & 94\% & 94\% \\
0.7 & 0.35 & 0.00  & 0.04 & 0.04 & 94\% & 94\% && 0.7 & 0.26 & 0.00  & 0.02 & 0.02 & 94\% & 94\% \\
0.8 & 0.35 & 0.00  & 0.05 & 0.05 & 93\% & 93\% && 0.8 & 0.25 & -0.00 & 0.03 & 0.03 & 94\% & 94\% \\
0.9 & 0.34 & -0.00 & 0.08 & 0.07 & 94\% & 94\% && 0.9 & 0.25 & -0.00 & 0.03 & 0.03 & 93\% & 93\%\\
\cline{1-7} \cline{9-15}
\end{tabular}
\end{center}

\newpage{}

\begin{center}
{\bf Appendix Table 4}\\
{\it Simulation results for $m=300$ and $n_i=10$ for all $i$, where:
$\alpha$ denotes the allocation probabilities; 
 $\mu_a$ denotes denotes the value of the target parameters $\mu(100,a,\alpha)$ for $a=0,1$;
Bias is the average of $\mu(100, a, \alpha) - \hat{\mu}(100, a, \alpha)$ for $a = 0,1$; 
ESE is the empirical standard error; ASE is the average of the sandwich variance based standard error estimates; 
EC denotes the empirical coverage of the $95\%$ Wald confidence intervals based on the Normal distribution;
and EC$_t$ denotes empirical coverage of $t$ distribution-based Wald CIs.}\\
\vspace{.1in}
\begin{tabular}{ccccccccccccccc}
\cline{1-7} \cline{9-15}
$\alpha$ & $\mu_0$ & Bias & ESE & ASE  & EC &EC$_t$ & & $\alpha$ & $\mu_1$ & Bias & ESE & ASE & EC &EC$_t$ \\ 
\cline{1-7} \cline{9-15}
0.1 & 0.39 & -0.02 & 0.10 & 0.09 & 90\% & 90\% && 0.1 & 0.28 & -0.01 & 0.10 & 0.09 & 88\% & 88\% \\
0.2 & 0.38 & -0.01 & 0.05 & 0.05 & 94\% & 94\% && 0.2 & 0.27 & -0.00 & 0.06 & 0.05 & 93\% & 93\% \\
0.3 & 0.37 & -0.01 & 0.04 & 0.04 & 95\% & 96\% && 0.3 & 0.27 & 0.00  & 0.04 & 0.04 & 94\% & 94\% \\
0.4 & 0.37 & -0.00 & 0.03 & 0.03 & 94\% & 94\% && 0.4 & 0.27 & 0.00  & 0.03 & 0.03 & 93\% & 93\% \\
0.5 & 0.36 & 0.00  & 0.03 & 0.03 & 94\% & 94\% && 0.5 & 0.26 & 0.00  & 0.02 & 0.02 & 93\% & 93\% \\
0.6 & 0.36 & 0.00  & 0.03 & 0.03 & 93\% & 93\% && 0.6 & 0.26 & 0.00  & 0.02 & 0.02 & 94\% & 94\% \\
0.7 & 0.35 & 0.00  & 0.03 & 0.03 & 92\% & 92\% && 0.7 & 0.26 & 0.00  & 0.02 & 0.02 & 93\% & 93\% \\
0.8 & 0.35 & 0.00  & 0.04 & 0.04 & 93\% & 93\% && 0.8 & 0.25 & -0.00 & 0.02 & 0.02 & 94\% & 94\% \\
0.9 & 0.34 & -0.01 & 0.06 & 0.06 & 94\% & 94\% && 0.9 & 0.25 & -0.00 & 0.03 & 0.03 & 93\% & 93\%\\
\cline{1-7} \cline{9-15}
\end{tabular}
\end{center}

\newpage{}

\begin{center}
{\bf Appendix Table 5}\\
{\it Simulation results for $m=400$ and $n_i=10$ for all $i$, where:
$\alpha$ denotes the allocation probabilities; 
 $\mu_a$ denotes denotes the value of the target parameters $\mu(100,a,\alpha)$ for $a=0,1$;
Bias is the average of $\mu(100, a, \alpha) - \hat{\mu}(100, a, \alpha)$ for $a = 0,1$; 
ESE is the empirical standard error; ASE is the average of the sandwich variance based standard error estimates; 
EC denotes the empirical coverage of the $95\%$ Wald confidence intervals based on the Normal distribution;
and EC$_t$ denotes empirical coverage of $t$ distribution-based Wald CIs.}\\
\vspace{.1in}
\begin{tabular}{ccccccccccccccc}
\cline{1-7} \cline{9-15}
$\alpha$ & $\mu_0$ & Bias & ESE & ASE  & EC &EC$_t$ & & $\alpha$ & $\mu_1$ & Bias & ASE & ASE & EC &EC$_t$ \\ 
\cline{1-7} \cline{9-15}
0.1 & 0.39 & -0.02 & 0.09 & 0.08 & 92\% & 92\% && 0.1 & 0.28 & -0.01 & 0.09 & 0.08 & 88\% & 88\% \\
0.2 & 0.38 & -0.01 & 0.05 & 0.04 & 94\% & 94\% && 0.2 & 0.27 & -0.00 & 0.05 & 0.05 & 93\% & 93\% \\
0.3 & 0.37 & -0.00 & 0.03 & 0.03 & 94\% & 95\% && 0.3 & 0.27 & -0.00 & 0.03 & 0.03 & 94\% & 94\% \\
0.4 & 0.37 & 0.00  & 0.03 & 0.03 & 95\% & 95\% && 0.4 & 0.27 & 0.00  & 0.02 & 0.02 & 94\% & 94\% \\
0.5 & 0.36 & 0.00  & 0.03 & 0.02 & 94\% & 94\% && 0.5 & 0.26 & 0.00  & 0.02 & 0.02 & 93\% & 93\% \\
0.6 & 0.36 & 0.00  & 0.03 & 0.02 & 94\% & 94\% && 0.6 & 0.26 & 0.00  & 0.02 & 0.02 & 92\% & 92\% \\
0.7 & 0.35 & 0.00  & 0.03 & 0.03 & 94\% & 94\% && 0.7 & 0.26 & 0.00  & 0.02 & 0.02 & 93\% & 94\% \\
0.8 & 0.35 & 0.00  & 0.03 & 0.03 & 95\% & 95\% && 0.8 & 0.25 & -0.00 & 0.02 & 0.02 & 93\% & 94\% \\
0.9 & 0.34 & -0.00 & 0.05 & 0.05 & 94\% & 95\% && 0.9 & 0.25 & -0.00 & 0.02 & 0.02 & 95\% & 96\%\\
\cline{1-7} \cline{9-15}
\end{tabular}
\end{center}

\
\end{document}